\documentclass[useAMS,usenatbib,twocolumn]{mn2e}
\usepackage[dvips]{graphicx}
\usepackage{amsmath}
\usepackage{amsfonts}
\usepackage{amssymb}

\title[Efficient sampling of limb darkening coefficients]{Efficient, 
uninformative sampling of limb darkening coefficients for two-parameter laws}
\author[David M. Kipping]{David M. Kipping$^{1,2}$\thanks{E-mail:
dkipping@cfa.harvard.edu}\\
$^{1}$Harvard-Smithsonian Center for Astrophysics, 60 Garden St., Cambridge, MA 
02138, USA \\
$^{2}$Carl Sagan Fellow}
\begin{document}

\date{Accepted 2013 July 31. Received 2013 July 30; in original form 2013 June 28}

\pagerange{\pageref{firstpage}--\pageref{lastpage}} \pubyear{2013}

\maketitle

\label{firstpage}

\begin{abstract}

Stellar limb darkening affects a wide range of astronomical measurements and
is frequently modelled with a parametric model using polynomials in the
cosine of the angle between the line of sight and the emergent intensity.
Two-parameter laws are particularly popular for cases where one wishes to fit
freely for the limb darkening coefficients (i.e. an uninformative prior) due to
the compact prior volume and the fact that more complex models rarely obtain 
unique solutions with present data. In such cases, we show that the two limb 
darkening coefficients are constrained by three physical boundary conditions, 
describing a triangular region in the two-dimensional parameter space. We show 
that uniformly distributed samples may be drawn from this region with optimal 
efficiency by a technique developed by computer graphical programming: 
triangular sampling. Alternatively, one can make draws using a uniform, 
bivariate Dirichlet distribution. We provide simple expressions for these 
parametrizations for both techniques applied to the case of quadratic, 
square-root and logarithmic limb darkening laws. For example, in the case of the 
popular quadratic law, we advocate fitting for $q_1 \equiv (u_1+u_2)^2$ and 
$q_2 \equiv 0.5u_1(u_1+u_2)^{-1}$ with uniform priors in the interval $[0,1]$ to 
implement triangular sampling easily. Employing these parametrizations allows 
one to derive model parameters which fully account for our ignorance about the 
intensity profile, yet never explore unphysical solutions, yielding robust and 
realistic uncertainty estimates. Furthermore, in the case of triangular sampling 
with the quadratic law, our parametrization leads to significantly reduced 
mutual correlations and provides an alternative geometric explanation as to why 
naively fitting the quadratic limb darkening coefficients precipitates 
strong correlations in the first place.

\end{abstract}


\section{Introduction}
\label{sec:intro}

Stellar limb darkening is the wavelength-dependent diminishing of the surface
brightness from the centre of the disc to the limb of the star. Limb darkening 
affects a wide range of different astronomical observations, such as optical 
interferometry (e.g. \citealt{aufdenberg:1995}), microlensing light curves 
(e.g. \citealt{witt:1995}; \citealt{zub:2011}), rotational modulations 
\citep{macula:2012}, eclipsing binaries \citep{kopal:1950} and transiting
planets \citep{mandel:2002}. Due to the often subtle, profile distorting
effects of limb darkening, the parameters describing limb darkening are 
frequently degenerate with other model parameters of interest, and thus accurate 
modelling is crucial in the interpretation of such data. 

Many of these astronomical phenomena may be described with precise closed-form 
analytic solutions, if one assumes a parametric limb darkening law. For example, 
the transit light curve may be expressed using hypergeometric functions and 
elliptical integrals when one adopts a polynomial law \citep{mandel:2002,
kjurkchieva:2013}. Such closed forms are not only computationally expedient to 
evaluate, but their parametrization also easily allows for uninformative priors 
on a target star's properties and Bayesian model selection of different laws, 
since the prior volume can be directly controlled.

Many of the commonly employed parametric limb darkening laws have been chosen to 
provide the best approximation possible between stellar atmosphere model 
intensity profiles and simple polynomial expansions (e.g. \citealt{claret:2000};
\citealt{claret:2003}; \citealt{sing:2010}; \citealt{hayek:2012}). This is 
because a typical approach was to regress a model to some observations whilst 
assuming a fixed stellar limb darkening law which most realistically described 
the modeller's expectation for the star. The benefits of this approach are that 
the parameters describing the limb darkening do not have to be varied, making 
the regressions considerably easier. However, an obvious consequence of this is 
that any model parameters derived from such an approach are fundamentally 
dependent upon the stellar atmosphere model adopted. An equivalent way of 
describing this approach is that a Dirac delta function prior was adopted for 
the limb darkening profile, which is statistically an implausible scenario.

An alternative strategy is to relax the constraint to weaker or even
uninformative priors. However, the trade-off is that by adopting a finite prior 
volume for the parameters describing the limb darkening, it is strongly 
preferred to use as compact a parametric model as possible (i.e. fewer 
parameters) so that the regression algorithm can reasonably hope to explore the 
full parameter space. Nevertheless, this is a statistically more robust approach 
than simply fixing these parameters, which are frequently correlated to the 
other model terms \citep{pal:2008}.

An example of a weaker prior would be to regress a joint probability density 
function (PDF) to the limb darkening coefficients (LDCs) emerging from an 
ensemble of stellar atmosphere models (e.g. \citealt{kepler22:2013}). However, 
even this approach is still fundamentally dependent upon stellar atmosphere 
models, since it is from these models that the ensemble of coefficients is 
initially computed. In contrast, uninformative priors make no assumption about 
the limb darkening profile, except for the parametric form which describes the 
intensity profile (e.g. the polynomial orders used). Such an approach may even 
be used to reverse engineer properties of individual stars or populations 
thereof \citep{neilson:2012}, although \citet{howarth:2011} cautions that one 
must carefully account for the system geometry when comparing fitted LDCs and 
those from stellar atmosphere models.

Adopting a simple parametric limb darkening law with uninformative priors is
therefore a powerful way of (i) incorporating and propagating our ignorance 
about the target star's true intensity profile into the derivation of all model 
parameters, (ii) presenting results which are independent of theoretical stellar 
atmosphere models, (iii) modelling astronomical phenomenon using closed-form and 
thus highly expedient algorithms and (iv) providing insights and constraints on 
the fundamental properties of the target star. 

The most common choice of uninformative prior for LDCs is a simple uniform 
prior. One danger of uninformative priors is that allowing the LDCs to explore 
any parameter range can often lead to unphysical limb darkening profiles being 
explored. It is therefore necessary to impose boundary conditions which prevent 
such violations. In this work, we show that after imposing the said boundary 
conditions (\S\ref{sub:derivation}), the PDF describing an uninformative joint 
prior on the quadratic LDCs is a uniform, bivariate Dirichlet distribution 
(\S\ref{sub:dirichlet}). Furthermore, we show that one may efficiently sample 
from this distribution using a trick from the field of computer graphical 
programming: triangular sampling (\S\ref{sub:triangular}). This results in a new 
parametrization for the quadratic LDCs which samples the physically plausible 
range of LDCs in an optimally efficient and complete manner. By comparing our 
results to previously proposed parametrizations, we show that this approach is 
at least twice as efficient as all others (\S\ref{sec:comp}). Finally, we 
provide optimal parametrizations using triangular sampling for other 
two-parameter limb darkening laws (\S\ref{sec:otherlaws}).

\section{Quadratic Limb Darkening Law}
\label{sec:quadratic}

\subsection{Deriving the three boundary conditions}
\label{sub:derivation}

We begin by considering the quadratic limb darkening law due its wide ranging 
use in a variety of fields. We first derive the boundary conditions which 
constrain the physically plausible range of the associated LDCs. Note, this is 
not the first presentation of such constraints (e.g. \citealt{burke:2007}), but 
due to some distinct constraints present elsewhere in the literature (e.g. 
\citealt{carter:2009}) and the fact that this derivation serves as a template 
for the applying constraints to other two-parameter limb darkening laws (e.g. 
see later \S\ref{sec:otherlaws}), we present an explicit derivation here.

We begin by considering the widely used quadratic limb darkening law. The
quadratic law seems to have first appeared in \citet{kopal:1950} and is
attractive due to its simple, intuitive form, flexibility to explore a range
of profiles plus a fairly compact, efficient structure. The specific intensity 
of a star, $I(\mu)$, following the quadratic limb darkening may be described by

\begin{align}
I(\mu)/I(1) &= 1 - u_1 (1-\mu) - u_2 (1-\mu)^2,
\label{eqn:Ispecific}
\end{align}

where $I(1)$ is the specific intensity at the centre of the disc, $u_1$ and
$u_2$ are the quadratic LDCs and $\mu$ is the cosine of the angle between the 
line of sight and the emergent intensity. We may also express 
$\mu=\sqrt{1-r^2}$, where $r$ is the normalized radial coordinate on the disc of 
the star.

We wish to investigate whether imposing some physical conditions on this 
expression leads to any useful constraints on the allowed ranges of the 
coefficients $u_1$ and $u_2$. In what follows, we define physically plausible
limb darkening profiles in reference to broad bandpass photometric/imaging 
observations of normal main-sequence stars (i.e. we do not consider pulsars,
white dwarfs, brown dwarfs, etc). Accordingly, we may impose the following two 
physical conditions:

\begin{itemize}
\item[{\textbf{(A)}}] an everywhere-positive intensity profile,
\item[{\textbf{(B)}}] a monotonically decreasing intensity profile from the
centre of the star to the limb.
\end{itemize}

Condition \textbf{(A)} requires little justification since a negative intensity
has no physical meaning and it may be expressed algebraically as 
$I(\mu)>0$ $\forall$ $0\leq\mu<1$, or

\begin{align}
u_1 (1-\mu) + u_2 (1-\mu)^2 < 1 \,\,\,\,\,\forall\,\,\,\,\,\,0\leq\mu<1.
\end{align}

The above can be evaluated in one of two extrema; minimizing the LHS with 
respect to $\mu$ and maximizing the LHS with respect to $\mu$. Consider 
first minimizing the LHS, which is trivially found to occur for 
$\mu\rightarrow1$. This leaves us with the meaningless constraint that $0<1$, 
which is of course satisfied for all $u_1$ and $u_2$ and thus leads to no useful 
constraints on the LDCs.

The other extrema of this condition is found by evaluating the LHS at its 
maximum, which is again trivially found to occur when $\mu\rightarrow0$ and 
leads us to

\begin{align}
u_1+u_2 < 1.
\label{eqn:conditionA}
\end{align}

Therefore, the physical requirement of an everywhere-positive intensity profile 
leads to a single constraint on the LDCs, given by 
equation~(\ref{eqn:conditionA}).

Next, let us enforce condition \textbf{(B)}, that the specific intensity is a 
monotonically decreasing function towards the limb. This is generally expected
for any broad bandpass limb darkening profile \citep{burke:2007}, but some 
narrow spectral lines, such as Si IV, could produce limb-brightened profiles 
\citep{schlawin:2010}. Focusing on the much more common case of limb darkening
though, we have

\begin{align}
\frac{ \partial I(\mu) }{\partial \mu } > 0,
\end{align}

which is easily shown to give

\begin{align}
u_1 + 2u_2(1-\mu) > 0.
\end{align}

One of the extrema of this condition is found by minimizing the LHS with respect 
to $\mu$, which occurs for $\mu\rightarrow1$, giving

\begin{align}
u_1 > 0.
\label{eqn:conditionB}
\end{align}

The other extrema occurs when we maximize the LHS with respect to $\mu$,
which occurs for $\mu\rightarrow0$ and gives

\begin{align}
u_1 + 2 u_2 > 0.
\label{eqn:conditionC}
\end{align}

We therefore derive two constraints on the LDCs from condition \textbf{(B)} 
(equations~\ref{eqn:conditionB} \& \ref{eqn:conditionC}). In total then, we have 
three boundary conditions on the coefficients $u_1$ and $u_2$:

\begin{align}
u_1+u_2 &< 1,\nonumber \\
u_1 &> 0,\nonumber \\
u_1 + 2 u_2 &> 0.
\label{eqn:conditions}
\end{align}

\subsection{Comparison to previous works}
\label{sub:theorycomparison}

Before proceeding to our new parametrization model, we pause to compare
our derived boundary conditions to those in previous works.
The first explicit declaration of a set of expressions used to enforce
physically plausible LDCs, that we are aware of, seems to come from 
\citet{burke:2007}. Here, the authors state all three of the same boundary 
conditions (see \S3.2 of that work) stated here in 
equation~(\ref{eqn:conditions}). This is not surprising as \citet{burke:2007}
enforced the same physical criteria [i.e. conditions \textbf{(A)} and 
\textbf{(B)}] to derive their expressions, i.e. an everywhere-positive intensity 
profile and a monotonically decreasing brightness from the centre-to-limb.

Another paper stating boundary conditions on the LDCs comes from 
\citet{carter:2009}, where the authors used the conditions $(u_1+u_2)<1$, 
$u_1>0$ and $(u_1+u_2)>0$. We point out that the last constraint
seems to be a typographical error missing a `2', but otherwise are the same
as those constraints provided here. We highlight this minor point to avoid
potential confusion in comparing these works.

\subsection{Visualizing the constraints}
\label{sub:visualizing}

\begin{figure}
\begin{center}
\includegraphics[width=8.4 cm]{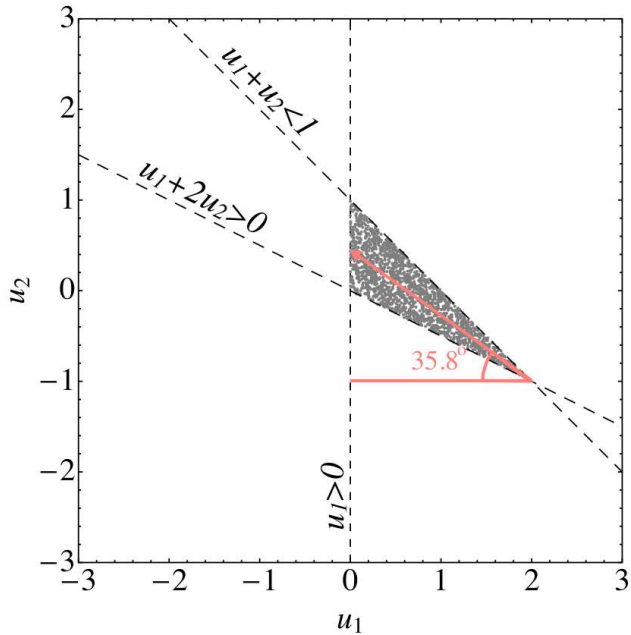}
\caption{\emph{Drawing $u_1$ and $u_2$ from a uniform distribution between
$-3$ and $+3$, we show the realizations which satisfy the physical constraints
of equation~(\ref{eqn:conditions}). The black dashed lines describe the three
constraints. The loci of accepted points form a triangle with a bisector
inclined $35.8^{\circ}$ to the $u_1$-axis.}} 
\label{fig:constraints}
\end{center}
\end{figure}

In order to visualize the constraints of equation~(\ref{eqn:conditions}), we
generated $u_1$ and $u_2$ by naively randomly sampling a uniform 
distribution bounded by $-3<u_1<+3$ and $-3<u_2<+3$. For every realization, we 
only accept the draw if all of the constraints in 
equation~(\ref{eqn:conditions}) are satisfied, as shown in 
Fig.~\ref{fig:constraints}. Iterating until $10^5$ trials were accepted, we 
required 3.6 million trials, i.e. an efficiency of 2.8\%. This highlights how 
inefficient it would be to sample from such a joint distribution.

\begin{figure}
\begin{center}
\includegraphics[width=8.4 cm]{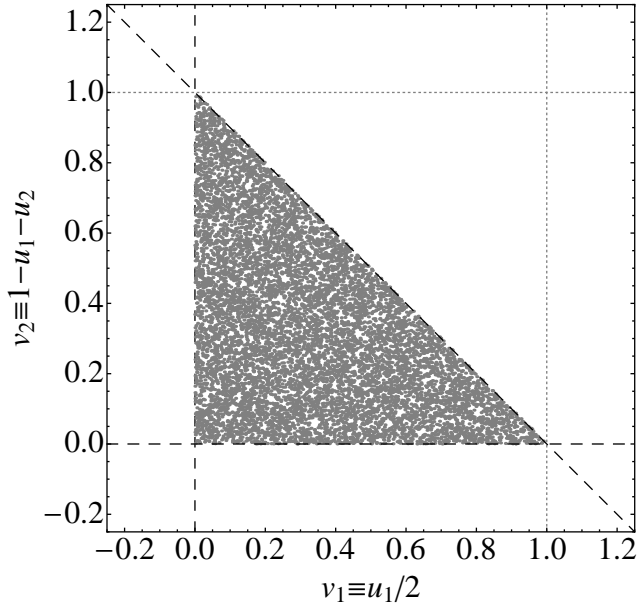}
\caption{\emph{Same as Fig.~\ref{fig:constraints}, except that we have
re-parametrized the two axes. One can see that the allowed physical range falls 
within a triangle which covers exactly one half of the unit square 
$\{0,0\}\rightarrow\{0,1\}\rightarrow\{1,1\}\rightarrow\{1,0\}$. This square 
describes the constraints stated in \citet{carter:2009}, which violate condition
\textbf{(B)}.
}} 
\label{fig:triangle}
\end{center}
\end{figure}

One may re-plot Fig.~\ref{fig:constraints} using different axes to visualize 
the constraints in alternative ways. We found that one particularly useful way 
of visualizing the constraints was found by plotting $v_1$ against $v_2$, as 
shown in Fig.~\ref{fig:triangle}, where we use the parametrization:

\begin{align}
v_1 &\equiv u_1/2,\\
v_2 &\equiv 1 - u_1 - u_2.
\end{align}

Using this parametrization, Fig.~\ref{fig:triangle} reveals the loci of points
satisfying conditions \textbf{(A)} and \textbf{(B}) form a right-angled triangle.

\subsection{Physical priors using the Dirichlet distribution}
\label{sub:dirichlet}

For those familiar with the Dirichlet distribution, the shape of 
Fig.~\ref{fig:triangle} will bear an uncanny resemblance to
the uniform, bivariate Dirichlet distribution. The Dirichlet distribution is a
multivariate generalization of the Beta distribution (which itself has useful
applications as a prior; \citealt{beta:2013}). Aside from being able to
exactly reproduce the distribution shown in Fig.~\ref{fig:triangle},
the bivariate Dirichlet distribution is able to reproduce a diverse range 
of profiles with just three so-called `concentration' parameters 
($\balpha=\{\alpha_1,\alpha_2,\alpha_3\}^T$). The PDF is given by

\begin{align}
\mathrm{P}(\balpha;v_1,v_2) &=\frac{v_1^{\alpha_2-1} v_2^{\alpha_1-1} (1-v_1-v_2)^{\alpha_3-1} \Gamma[\alpha_1+\alpha_2+\alpha_3]}
{\Gamma[\alpha_1] \Gamma[\alpha_2] \Gamma[\alpha_3]},
\label{eqn:dirichlet}
\end{align}

for $v_1 > 0$, $v_2 > 0$ and  $(v_1+v_2)<1$; otherwise 
$\mathrm{P}(\balpha;v_1,v_2) = 0$. In the case of the uniform distribution 
of Fig.~\ref{fig:triangle}, one may simply use $\balpha = \mathbf{1}$:

\begin{equation*}
\mathrm{P}(\balpha=\mathbf{1};v_1,v_2) =
\begin{cases}
2  & \text{if} \ v_1>0 \ \mathrm{\&} \ v_2>0 \ \mathrm{\&} \ (v_1+v_2)<1 ,\\
0 & \text{otherwise}.
\end{cases}
\label{eqn:flatdirichlet}
\end{equation*}

The bivariate Dirichlet distribution is also uniquely defined over the 
range $v_1>0$, $v_2>0$ and $(v_1+v_2)<1$ and naturally integrates to unity
over this range. It may therefore be used to serve as a proper prior.

\subsection{Physical priors using triangular sampling}
\label{sub:triangular}

Consider the special case where one requires sampling from a uniform prior
in the joint distribution $\{u_1,u_2\}$ (but wishes to enforce that all
sampled realizations are physical). This corresponds to the uniform, bivariate
Dirichlet distribution described by equation~(\ref{eqn:dirichlet}) with
$\balpha=\mathbf{0}$. One therefore needs to simply draw a random variate in 
$\{v_1,v_2\}$ from the uniform, bivariate Dirichlet distribution. 

However, another way of thinking about the problem is to try to populate a 
triangle with a uniform sampling of points, as evident from 
Fig.~\ref{fig:triangle} - a procedure we dub `triangular sampling'. This more
geometric perspective leads to a simple and elegant expression for generating 
the LDC samples. 

An elegant method for triangular sampling comes from the field of computer
graphical programming, which we will describe here. Consider a triangle with 
vertices $\mathbf{A}$, $\mathbf{B}$ and $\mathbf{C}$, and two random uniform 
variates $q_1$ and $q_2$ in the interval $[0,1]$. \citet{turk:1990} showed that 
a random location, $\mathbf{v}$, within the triangle can be sampled using
(notation has been changed slightly from that of \citealt{turk:1990})

\begin{align}
\mathbf{v} &= (1-\sqrt{q_1}) \mathbf{A} + \sqrt{q_1} (1-q_2) \mathbf{B} 
+ q_2 \sqrt{q_1} \mathbf{C}.
\end{align}

\begin{figure}
\begin{center}
\includegraphics[width=8.4 cm]{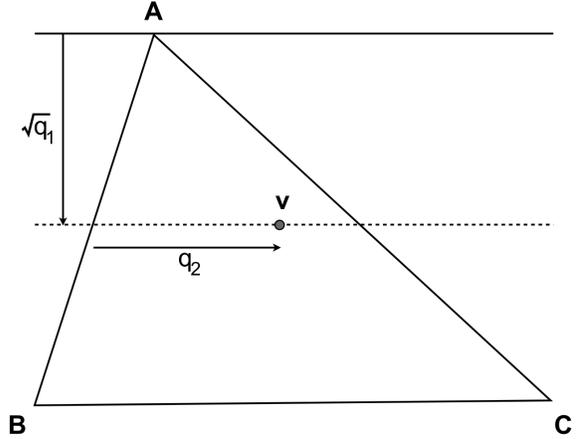}
\caption{\emph{A geometric illustration of how a random point is drawn
from a triangle with vertices $A$, $B$ and $C$ using two random variates
$q_1$ and $q_2$ (i.e. `triangular sampling'). The method and figure adapted 
are from the computer graphical programming chapter of \citet{turk:1990}.
}} 
\label{fig:turk}
\end{center}
\end{figure}

This sampling is equivalent to having $q_1$ draw out a line segment parallel to
$BC$ that joins a point on $AB$ with a point on $AC$ and then selecting a point
on this segment based upon the value of $q_2$ (as shown in 
Fig.~\ref{fig:turk}). Taking the square root of $q_1$ is necessary to weight
all portions of the triangle equally.

Evaluating the above for $\mathbf{A}=\{0,1\}^T$, $\mathbf{B}=\{0,0\}^T$ and
$\mathbf{C}=\{1,0\}^T$ (representing the vertices of the specific triangle we 
are interested in) gives

\begin{align}
v_1 &= \sqrt{q_1} q_2,\\
v_2 &= 1-\sqrt{q_1}.
\end{align}

Substituting the above into equations (9) and (10) gives

\begin{align}
u_1 &= 2 \sqrt{q_1} q_2,\\
u_2 &= \sqrt{q_1} (1 - 2 q_2).
\label{eqn:myeqn}
\end{align}

The inverse of these expressions are easily found to be given by

\begin{align}
q_1 &\equiv (u_1 + u_2)^2,\\
q_2 &\equiv \frac{u_1}{2(u_1 + u_2)}.
\label{eqn:myinverseeqn}
\end{align}

By re-parametrizing the LDCs from a set $\btheta$ of $\btheta=\{u_1,u_2\}$ to 
$\btheta=\{q_1,q_2\}$, one can fit for quadratic LDCs in such a way that the 
joint prior distribution is uniform and exclusively samples physically plausible 
solutions. For example, one would fit for $q_1$ and $q_2$ with uniform priors 
between $0$ and $1$, but convert these parameters into $u_1$ and $u_2$ (using 
equation~15 \& 16) before calling their light curve generation code e.g. 
the \citet{mandel:2002} algorithm. This will exactly reproduce the uniform, 
bivariate Dirichlet distribution shown in Fig.~\ref{fig:constraints} \& 
\ref{fig:triangle}.

\subsection{Comparison to theoretical stellar atmosphere models}
\label{sub:theorycomp}

By sampling from $\btheta=\{q_1,q_2\}$ uniformly over the interval $[0,1]$, one 
adopts uninformative priors in the LDCs and thus the underlying intensity 
profile of a given star. The only physics which goes into our model are the 
conditions \textbf{(A)} and \textbf{(B)}. In contrast, LDCs generated using 
stellar atmosphere models include a great deal of physics, and sampling 
coefficients from such a model is more appropriately described as using 
informative priors. The choice as to which path to follow is a matter for the 
data analyst to decide and is likely dependent upon how well characterized the 
target star is and how much trust is placed in the theoretical models.

An implicit expectation of our $\btheta=\{q_1,q_2\}$ model is that the true LDCs 
of normal stars observed in a broad bandpass should fall within the unit-square
of $0<q_1<1$ and $0<q_2<1$. By extension then, the LDCs of a realistic
stellar atmosphere should also reproduce coefficients lying within this
unit-square. To check this, we here show the results of converting standard 
tabulations of quadratic LDCs into the $\btheta=\{q_1,q_2\}$ parametrization. We
decide to use the \emph{Kepler} bandpass for this comparison since our model is 
(a) designed for broad bandpass photometry, (b) most useful for faint target 
stars with poor characterization requiring uninformative priors and (c) likely 
to be most commonly employed on such targets due to the sheer volume of 
observations obtained by this type of survey.

\citet{claret:2011} provide tabulations of \emph{Kepler} LDCs for the quadratic 
law computed using 1D Kurucz ATLAS\footnote{http://kurucz.harvard.edu/} and 
PHOENIX\footnote{http://www.hs.uni-hamburg.de/EN/For/ThA/phoenix/}
stellar atmosphere models over a wide range of stellar input
parameters: $0.0\leq\log g\leq5.0$, $-5\leq[\mathrm{M}/\mathrm{H}]\leq+1$,
$2000\leq T_{\mathrm{eff}}\leq50000$\,K. The extreme ends of this temperature
range do not necessarily conform to the criteria $\textbf{(A)}$ and 
$\textbf{(B)}$, even in \emph{Kepler's} broad bandpass, and so we make some
cuts to avoid the extrema. The lowest effective temperature for a planet-hosting 
star is Kepler-42 (aka KOI-961) with $T_{\mathrm{eff}}=3068\pm174$\,K 
\citep{muirhead:2012} and so we make a cut at 3000\,K. The highest effective 
temperature of a planet-hosting star is $8590\pm73$\,K for Formalhaut b 
\citep{kalas:2008}, and so we place an additional cut at $10000$\,K.

\begin{figure}
\begin{center}
\includegraphics[width=8.4 cm]{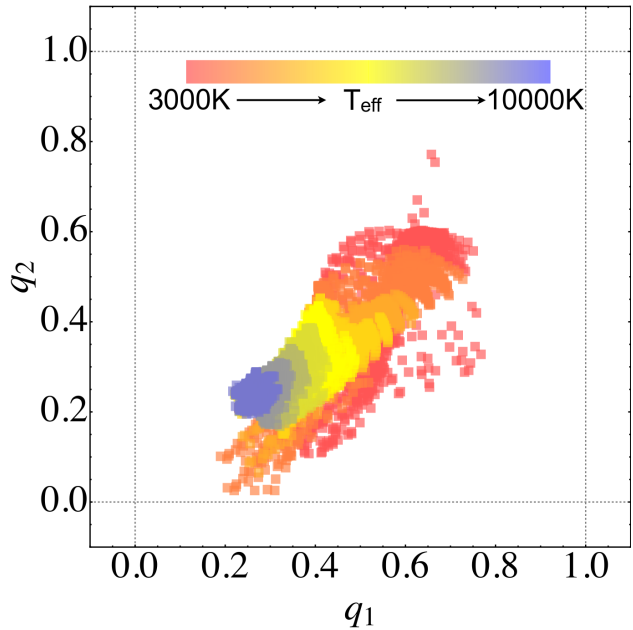}
\caption{\emph{Quadratic LDCs generated from stellar atmosphere models over the 
\emph{Kepler} bandpass by \citet{claret:2011}.The original LDCs ($u_1$-$u_2$) 
have been re-parametrized into our $q_1$-$q_2$ scheme. Stellar parameters range 
from $0\leq\log g\leq+5$, $-5\leq[\mathrm{M}/\mathrm{H}]\leq+1$ and 
$4000\leq T_{\mathrm{eff}}\leq10000$\,K, with the latter indicated by the colour
of the points (blue=hot; red=cool).}}
\label{fig:theorycomp}
\end{center}
\end{figure}

Using this range and the \citet{claret:2011} tabulations, we compute
$\{q_1,q_2\}$ from $\{u_1,u_2\}$ for all 12026 entries and display the results
in Fig.~\ref{fig:theorycomp}. It can be easily seen that the entire grid
falls within the expected unit-square. We therefore conclude that our 
parametrization is consistent with the results from a typical stellar atmosphere 
model. It is interesting to observe that hot-stars display a narrow range of
LDCs in the \emph{Kepler} bandpass since the Wien's peak wavelength is
sufficiently low that the Rayleigh tail dominates the part of the spectrum
seen by \emph{Kepler}.

\section{Comparison to Previously Suggested Parametrizations}
\label{sec:comp}

\subsection{Overview}
\label{sub:compoverview}

In order to give our proposed parametrization some context, we here discuss 
previously suggested parametrizations of the LDC, with sole attention given to 
the quadratic law \citep{kopal:1950}, due to its very frequent use, particularly 
in the transiting exoplanet community. There have been numerous distinct 
suggestions for reasonable parametrizations in the exoplanet literature, and 
here, we attempt to compare our proposed parametrization to those of the 
previous ones (that we are aware of at least).

\subsection{A comment on mutual correlations}
\label{sub:correls}

Before we continue, there is an important point we would like to establish.
Many of the previous suggestions have been designed to minimize the correlation 
between the two fitted limb darkening parameters whilst regressing data and 
\emph{not} specifically designed to sample the physically plausible solutions in 
an efficient and complete manner (which is the motivation behind our 
parametrization). So, which motivation is preferable?

In most cases, astronomical data do not usually constrain freely fitted LDCs 
particularly well. For example, for rotational modulations the limb darkening 
profile is degenerate with the spots' contrasts and geometries 
\citep{macula:2012}. In the case of transiting planets or eclipsing binaries, 
the LDCs are degenerate with the geometry and size of the eclipsing body 
\citep{pal:2008,howarth:2011}. Therefore, in most cases, the data are 
essentially unconstraining and we do not improve our ignorance of the true 
profile significantly. The power of our technique lies in the fact that by 
efficiently sampling the entire physically plausible parameter volume, we 
propagate that ignorance into the posterior distributions of all of the 
parameters which are correlated to the LDCs. So by fitting for $\{q_1,q_2\}$ 
with uniform priors over the interval $[0,1]$, the derived posteriors account 
for the full range of physically permissible models.

Additionally, the only consequence of fitting two parameters with non-zero
mutual correlation is that more computational resources are required to obtain
a converged solution e.g. for Markov Chain Monte Carlo (MCMC) this would require 
a greater chain length. However, this issue is actually somewhat less important 
in the modern age of computing with significant strides in CPU speeds. We 
therefore argue that it is more valuable to sample from a physically plausible 
prior volume.

Finally, it is important to realize that despite the very wide use of MCMC
techniques, other regression techniques are becoming increasingly popular and
have different issues affecting their efficiency. Suppose a set of data strongly 
constrains the quadratic LDCs. For MCMC \citep{metropolis:1953,hastings:1970}, 
one could seed the chain from the approximate solution location and because the 
data is constraining, the chain should never cross the three boundary conditions 
(i.e. it is highly efficient). In contrast, for nested sampling 
\citep{skilling:2004}, the initial nests stretch across the entire prior volume 
and thus any regions which violate the boundary conditions would have to be 
rejected through a likelihood penalty, and thus the larger this region is, the 
poorer the efficiency of the nested sampling algorithm. As a side note, in the 
case of poorly constraining data, nested sampling is the more efficient code 
since the MCMC routine would randomly walk into unphysical territory frequently, 
but (with well-chosen priors) nested sampling will not.

Despite not being designed to minimize the mutual correlation between the
two fitted LDCs, numerical experiments show that our parametrization does in 
fact reduce the correlation significantly. In recent months, the Hunt for 
Exomoons with Kepler (HEK) project \citep{hek:2012} has been employing our 
proposed parametrization during their fits of \emph{Kepler} transiting planetary 
candidates, and initial results find that the mutual correlation is reduced from 
a median value of $\mathrm{Corr}[u_1,u_2]=-0.89$ to 
$\mathrm{Corr}[q_1,q_2]=-0.37$ (see Table~\ref{tab:correls}). The reason why our 
parametrization reduces the correlation can be explained geometrically and is 
discussed in \S\ref{sub:pal}.

We therefore argue that efficient, complete sampling of the physically plausible
prior volume has many advantages over simply reducing mutual correlation, which
is why we have pursued such an approach in this paper. However, a by-product of 
our proposition is that mutual correlations are significantly reduced anyway.

\begin{table*}
\caption{\emph{Mutual correlations of quadratic LDCs for two different 
parametrizations. Although it is not the purpose of the $\{q_1,q_2\}$ 
parametrization, the mutual correlation is significantly reduced. Fits comes 
from results in preparation by the HEK project, except HCO-254.01 which is 
Kepler-22b \citep{kepler22:2013}. Planet ($\mathcal{P}$) and satellite 
($\mathcal{S}$) model nomenclature follows \citet{kepler22:2013}.
}}
\centering 
\begin{tabular}{l l l l} 
\hline\hline 
HEK candidate ID & Model & $u_1$-$u_2$ Correlation & $q_1$-$q_2$ Correlation \\ [0.5ex] 
\hline
HCO-254.01	& $\mathcal{P}_{\mathrm{LD-free}}$	& $-0.946029$	& $-0.290086$ \\
HCO-254.01	& $\mathcal{P}_{\mathrm{LD-free},e_{B*}}$& $-0.878806$	& $-0.507617$ \\
HCO-254.01	& $\mathcal{S}_{\mathrm{LD-free}}$	& $-0.925289$	& $-0.211113$ \\
HCO-254.01	& $\mathcal{S}_{\mathrm{LD-free},e_{B*}}$& $-0.891628$	& $-0.406471$ \\
HCO-254.01	& $\mathcal{S}_{\mathrm{LD-free},e_{SB}}$& $-0.951899$	& $-0.269654$ \\
HCA-39.02	& $\mathcal{P}_{\mathrm{LD-free}}$	& $-0.931938$	& $+0.180139$ \\
HCA-39.02	& $\mathcal{S}_{\mathrm{LD-free}}$	& $-0.956056$	& $-0.374412$ \\
HCA-669.01	& $\mathcal{P}_{\mathrm{LD-free}}$	& $-0.949620$	& $-0.001698$ \\
HCA-669.01	& $\mathcal{S}_{\mathrm{LD-free}}$	& $-0.571816$	& $-0.480137$ \\
HCO-754.01	& $\mathcal{P}_{\mathrm{LD-free}}$	& $-0.713572$	& $-0.447555$ \\
HCO-754.01	& $\mathcal{S}_{\mathrm{LD-free}}$	& $-0.703587$	& $-0.172976$ \\
HCV-531.01	& $\mathcal{P}_{\mathrm{LD-free}}$	& $-0.580507$	& $-0.482708$ \\
HCV-531.01	& $\mathcal{S}_{\mathrm{LD-free}}$	& $-0.567252$	& $-0.472353$ \\
HCA-941.01	& $\mathcal{P}_{\mathrm{LD-free}}$	& $-0.599955$	& $-0.583683$ \\
HCA-941.01	& $\mathcal{S}_{\mathrm{LD-free}}$	& $-0.597042$	& $-0.575972$ \\
HCV-40.01	& $\mathcal{P}_{\mathrm{LD-free}}$	& $-0.986092$	& $-0.116053$ \\
HCV-40.01	& $\mathcal{S}_{\mathrm{LD-free}}$	& $-0.985540$	& $-0.104638$ \\ 
\hline
Median		& -					& $-0.891628$	& $-0.374412$ \\ [1ex]
\hline\hline 
\end{tabular}
\label{tab:correls} 
\end{table*}

\subsection{Performance metrics}
\label{sub:metrics}

Each parametrization has two metrics which describe how well they sample the
parameter space. We denote ``efficiency'', $\epsilon$, as unity minus the
fraction of times the parametrization produces an unphysical intensity profile
(which would require rejection). In a practical case, unphysical trials would 
have to be rejected in a Monte Carlo fit and thus act to reduce the overall 
efficiency and hence the name for this term. This value is easily calculated 
with a Monte Carlo experiment of $N\gg1$ synthetic draws from a given joint 
distribution.

The other metric we consider is `completeness', $\kappa$, which describes what 
fraction of the allowed physical parameter space is explored by the 
parametrization. A $\kappa<1$ means that certain regions of reasonable and 
physically plausible realizations of $\{u_1,u_2\}$ are never explored. The 
$\kappa$ value of a given parametrization, $\btheta$, is simply equal to the 
area of the loci sampled in $\{u_1,u_2\}$ parameter space divided by that 
achieved for only the physically acceptable trials (which happens to equal 
unity) multiplied by the efficiency, $\epsilon$. Note that the parametrization 
described in this work ($\btheta=\{q_1,q_2\}$; equation~17 \& 18) produces 
$\kappa=1$ and $\epsilon=1$, by virtue of its construction.

As mentioned in \S\ref{sub:correls}, one could argue that the correlation 
between the two fitted limb darkening parameters is also a key metric of 
interest. However, we make the case here that correlation is not critical in 
light of the substantial improvements in computational hardware and software 
over the last decade.

We will therefore proceed by considering several popular parametrizations of
the LDCs (in chronological sequence) and evaluating their efficiency, 
$\epsilon$, and completeness, $\kappa$. During this investigation, we found that 
it is quite rare for authors to declare the upper and lower bounds used on their 
priors (which are usually uniform). Without the prior bounds, it is not possible 
to evaluate $\epsilon$ and $\kappa$ and so in these cases we proceed by 
selecting a choice of prior bounds which ensures $\kappa=1$ for the highest 
possible $\epsilon$.

\subsection{$\btheta=\{u_1,u_2\}$}
\label{sub:simple}

We begin by first considering the naive parameter set of 
$\btheta=\{u_1,u_2\}$ directly, which serves as a useful baseline for subsequent 
comparisons. In order to estimate $\kappa$ and $\epsilon$ though, we must first 
choose upper/lower bounds on these two parameters. As discussed in the previous 
subsection, we can optimize the prior bounds to ensure $\kappa=1$. This is done 
by generating $N\gg1$ Monte Carlo realizations of $\{u_1,u_2\}$ across an overly
generous interval (in this case we used $[-3,+3]$) and only accepting points 
which satisfy the conditions stipulated in equation~(\ref{eqn:conditions}). We 
find that using $0<u_1<+2$ and $-1<u_2<+1$ ensures $\kappa=1$ for the highest 
possible efficiency. The corresponding value of $\epsilon$ is 0.25 i.e. sampling 
from this prior with a lack of constraining data would mean that three out of 
four trials would have to be rejected. The parameter volume sampled by this 
prior is illustrated in Fig.~\ref{fig:comps}.

\subsection{$\btheta=\{u_{+},u_{-}\}\equiv\{u_1+u_2,u_1-u_2\}$}
\label{sub:brown}

The pioneering work of \citet{brown:2001} offers perhaps the first such example 
of serious consideration of alternative parametrizations in the exoplanet
literature. Using the \emph{Hubble Space Telescope} photometry of HD~209458b, 
\citet{brown:2001} suggested fitting for $u_{+} \equiv (u_1 + u_2)$ and $u_{-} 
\equiv (u_1 - u_2)$. Unlike in this work, the purpose of this parametrization 
was not to ensure physically plausible intensity profiles, but rather to reduce 
the correlation between the two LDCs in the fitting procedure, as stated in 
\S3.2 of \citet{brown:2001}. 

In order to compute our performance metrics, a choice of prior bounds is 
required. We choose to select these bounds such that we optimize $\kappa=1$,
as discussed earlier in \S\ref{sub:metrics}. Following the same method described
in \S\ref{sub:simple}, we estimate that this occurs for $0<u_{+}<+1$ and 
$-1<u_{-}<+3$. Using these bounds, we calculate that $\epsilon=0.5$. This can
be intuitively visualized in Fig.~\ref{fig:comps}.

\subsection{$\btheta=\{U_1,U_2\}\equiv\{2u_1+u_2,u_1-2u_2\}$}
\label{sub:holman}

\citet{holman:2006} chose to fit for $U_1 \equiv (2u_1+u_2)$ and $U_2 \equiv 
(u_1-2u_2)$ because ``the resulting uncertainties in those parameters are 
uncorrelated''. Once again then, it is worth noting that the motivation of this 
parameter set was not to sample the physically allowed parameter space 
efficiently. The priors used in the exploration of these parameters are not 
stated in the paper, and so we assume uniform between some upper and lower 
bounds on each term. The numerical range is not stated in \citet{holman:2006} 
but we have learned that the exploration was unbound, but with rejections 
applied to samples which fall outside of the conditions stated in 
equation~(\ref{eqn:conditionA}), (\ref{eqn:conditionB}) \& 
(\ref{eqn:conditionC}) (private communication with M. Holman \& J. Winn).

We therefore proceed by optimizing the prior bound choice to $\kappa=1$ via the
same Monte Carlo method described earlier (\S\ref{sub:simple}). This procedure 
yields $0<U_1<+3$ and $-2<U_2<+4$. Using these limits, we calculate 
$\epsilon=0.278$ for the fixed choice of $\kappa=1$, which is illustrated in
Fig.~\ref{fig:comps}.

\subsection{$\btheta=\{a_1,a_2\}\equiv\{u_1+2u_2,2u_1-u_2\}$}
\label{sub:burke}

During our analysis of the literature on this subject, we noticed that the
parametrization of \citet{holman:2006} was cited by many authors including the
instance of \citet{burke:2007}. What is interesting is that \citet{burke:2007}
state that `we follow \citet{holman:2006} by adopting $a_1\equiv(u_1+2u_2)$ and
$a_2\equiv(2u_1-u_2)$', but as discussed earlier \citet{holman:2006} in fact 
used $U_1\equiv(2u_1+u_2)$ and $U_2\equiv(u_1-2u_2)$. Therefore, despite 
\citet{burke:2007} claiming to have simply followed \citet{holman:2006}, they 
had in fact introduced an entirely new parametrization. We explore this 
parametrization here.

\citet{burke:2007} do explicitly declare that they use uniform priors on the
LDCs but do not explicitly state the bounds on $a_1$
and $a_2$. However, the authors do state they impose $u_1>0$, $(u_1+u_2)<1$
and $(u_1+2u_2)>0$, which are identical to the conditions derived in this work
(see equations~\ref{eqn:conditionA}, \ref{eqn:conditionB} \& 
\ref{eqn:conditionC}). The $a_1$ parameter is therefore bound by $a_1>0$ but the 
other constraints do not naturally impose any other bounds. We are also unable 
to find any way of inferring any other bounds from the paper of 
\citet{burke:2007}.

We therefore proceed by selecting bounds on $a_1$ and $a_2$ ourselves and we 
choose parameters which optimize to $\kappa=1$, as discussed in 
\S\ref{sub:metrics}. Following the same Monte Carlo method used previously (e.g.
see \S\ref{sub:simple}), we determine $0<a_1<2$ and $-1<a_2<+5$ to ensure 
$\kappa=1$. Utilizing these bounds, we estimate $\epsilon=(5/12)=0.417$, which
is visualized in Fig.~\ref{fig:comps}.

\subsection{$\btheta=\{w_1,w_2\}\equiv\{u_1\cos\phi-u_2\sin\phi,u_1\sin\phi+
u_2\cos\phi\}$}
\label{sub:pal}

\citet{pal:2008} proposed that the correlation between $u_1$ and $u_2$ can
be minimized by using principal component analysis. The author suggested
the parametrization $w_1 \equiv (u_1\cos\phi-u_2\sin\phi)$ and $w_2 \equiv 
(u_1\sin\phi + u_2\cos\phi)$, where $0<\phi<\pi/2$ and is chosen such that the 
correlation is minimized. Once again then, we stress that this parametrization 
is not designed to sample from the physically plausible parameter space. 
\citet{pal:2008} does not suggest bounds on $w_1$-$w_2$ and so we proceed to
select bounds in such way as to optimize $\kappa=1$. This optimization
process is sensitive to $\phi$ though and it is possible to derive different
bounds depending upon what one assumes for $\phi$.

In order to explore this issue fully, we fix $\phi$ to a specific value between
$0$ and $\pi/2$ and then optimize the bounds on $w_1$ and $w_2$ to ensure
$\kappa=1$, using the same Monte Carlo method employed earlier (e.g. see
\S\ref{sub:simple}). We then use these bounds to compute $\epsilon$ as usual. 
For each choice of $\phi$ then, we compute a unique value of $\epsilon$, i.e.
$\epsilon(\phi)$. Repeating over a wide range of $\phi$ values, we find that
$\phi=45^{\circ}$ yields the maximum efficiency of $\epsilon=0.5$ and drops
to $\epsilon=0.25$ as one rotates round to $\phi=0^{\circ}$ and 
$90^{\circ}$.

Setting $\phi=45^{\circ}$ then optimizes the efficiency of sampling the
physically plausible parameter space. We stress that this choice is not made
to minimize the correlation between $w_1$ and $w_2$, for which we 
note \citet{pal:2008} recommend $\phi=35^{\circ}$-$40^{\circ}$. For the 
$\phi=45^{\circ}$ case, however, $w_1=(u_1-u_2)=u_{-}$ and 
$w_2=(u_1+u_2)=u_{+}$, and so we recover the same parametrization used by 
\citet{brown:2001}. For this reason, we do not include the parametrization of 
\citet{pal:2008} in Fig.~\ref{fig:comps} and Table~\ref{tab:comps}.

One interesting point is that the boundary conditions in 
equation~(\ref{eqn:conditionA}) and equation~(\ref{eqn:conditionC}) form two 
sides of the triangle described in Fig.~\ref{fig:constraints} and taking the 
bisector of these two lines yields a line inclined by 
$\phi=\frac{1}{2}[\tan^{-1}(\frac{1}{2})+\tan^{-1}(\frac{2}{2})]=35.8^{\circ}$, 
which is also marked in Fig.~\ref{fig:constraints}. Therefore, the suggested 
angle of $\phi=35^{\circ}$-$40^{\circ}$ by \citet{pal:2008} effectively just 
travels up along this bisector. Indeed, one can consider this to be an 
alternative geometric explanation for the suggestion of \citet{pal:2008}. It 
also highlights how our parametrization, $\{q_1,q_2\}$, should be expected to 
exhibit inherently low mutual correlation since it also travels up along this 
bisector line. This was indeed verified to be the case earlier in 
\S\ref{sub:correls} and here we are able to provide the explicit explanation for 
this observation.

\subsection{$\btheta=\{u_1,u_{+}\}\equiv\{u_1,u_1+u_2\}$}
\label{sub:carter}

\begin{figure}
\begin{center}
\includegraphics[width=8.4 cm]{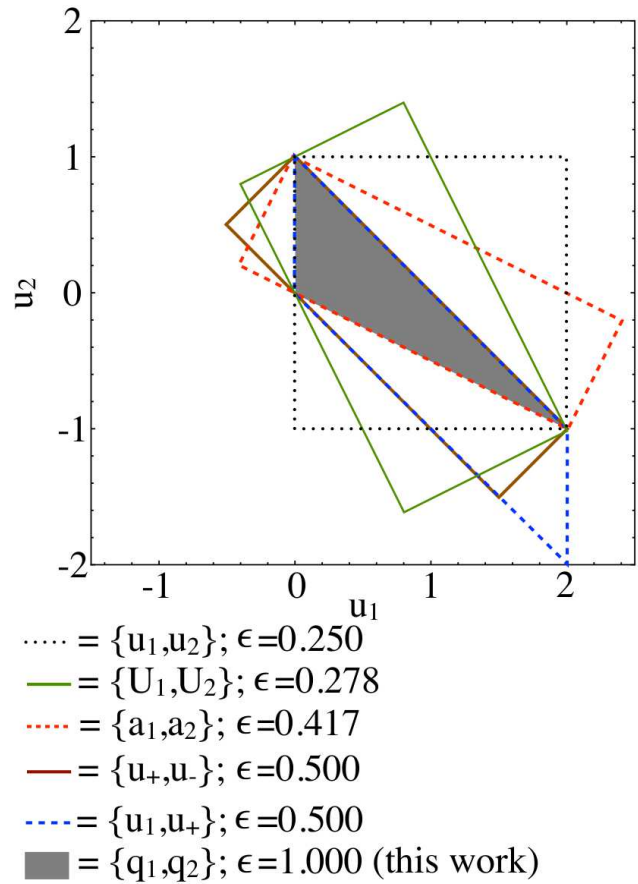}
\caption{\emph{Loci of points sampled by various parametrizations of the
quadratic LDCs. In each case, the completeness, $\kappa$, equals unity since we 
have optimized the prior bounds to ensure this condition. This is done since we 
are unable to find corresponding upper/lower bounds in the referenced 
literature. Grey area represents the physically plausible parameter range.
}} 
\label{fig:comps}
\end{center}
\end{figure}

The final parametrization we consider is that of 
$\btheta=\{u_1,u_{+}\}\equiv\{u_1,u_1+u_2\}$, which has been used in papers such
as \citet{nesvorny:2012} and \citet{hek:2013}. The choice of bounds here is
usually stated to be $0<u_1<+2$ and $0<(u_1+u_2)<+1$, which incidentally is
the same result that one finds when one optimizes the bounds to $\kappa=1$.
The loci of points form a parallelogram on the $\{u_1,u_2\}$ plane (as shown
in Fig.~\ref{fig:comps}), unlike any of the previously considered 
parametrizations which formed rectangles (or a triangle in the case of 
$\btheta=\{q_1,q_2\}$) and produces an efficiency of exactly one half i.e. 
$\epsilon=0.5$. Table~\ref{tab:comps} shows the efficiency and bounds of this
parametrization in relation to previously considered ones.

\begin{table*}
\caption{\emph{Comparison of the performance metric $\epsilon$ (efficiency) for
several different parametrizations of the quadratic LDCs. In each case, the 
completeness, $\kappa$, equals unity since we have optimized the prior bounds to 
ensure this condition. This is done since we are unable to find corresponding 
upper/lower bounds in the referenced literature.}} 
\centering 
\begin{tabular}{l c c c c} 
\hline\hline 
Parametrization, $\btheta$ & Parameter 1 Interval & Parameter 2 Interval & Efficiency, $\epsilon$  \\ [0.5ex] 
\hline 
$\{u_1,u_2\}$						& $[0,+2]$	& $[-1,+1]$ 	& $0.250$ \\ 
$\{u_{+},u_{-}\}\equiv\{u_1+u_2,u_1-u_2\}$			& $[0,+1]$	& $[-1,+3]$	& $0.500$ \\ 
$\{a_1,a_2\}\equiv\{2u_1+u_2,u_1-2u_2\}$			& $[0,+2]$	& $[-1,+5]$	& $0.417$ \\ 
$\{U_1,U_2\}\equiv\{u_1+2u_2,2u_1-u_2\}$			& $[0,+3]$	& $[-2,+4]$	& $0.278$ \\ 
$\{u_1,u_{+}\}\equiv\{u_1,u_1+u_2\}$				& $[0,+2]$	& $[0,+1]$	& $0.500$ \\ 
\hline
$\{q_1,q_2\}\equiv\{(u_1 + u_2)^2,(u_1/2)(u_1 + u_2)^{-1}\}$	& $[0,+1]$	& $[0,+1]$	& $1.000$ \\ [1ex] 
\hline\hline 
\end{tabular}
\label{tab:comps} 
\end{table*}

\section{Other Two-Parameter Limb Darkening Laws}
\label{sec:otherlaws}

\subsection{General principle}
\label{sub:general}

Although the quadratic limb darkening is the most widely used two-parameter
limb darkening in the literature, the so-called `square-root' law and to a
lesser extent the `logarithmic' law have also gained traction. Like the
quadratic case, enforcing the physical conditions of an everywhere-positive
intensity profile and a monotonically decreasing intensity from centre-to-limb
imposes three boundary conditions on the two coefficients describing each law.
Hence, we once again have a two-dimensional plane featuring three (non-parallel) 
boundary conditions which enclose a triangle. Therefore, sampling from this 
triangle in a uniform manner can be achieved using exactly the same trick 
described for the quadratic law case.

One can actually take this a step further and state that for \emph{any} problem 
with two variables with a uniformly distributed joint PDF and three 
(non-parallel) boundary conditions, 100\% complete and efficient sampling is 
easily achieved using the triangular sampling technique discussed in this paper.

\subsection{Square-root law}
\label{sub:sqrtlaw}

Arguably, the second-most popular two-parameter limb darkening law is that of
the square-root law. \citet{hamme:1993} argues that this is a superior 
approximation to the quadratic law for late-type stars in the near-infrared. 
Recent examples include applications to the eclipsing binary system 
LSPM J1112+7626 \citep{irwin:2011} and the transiting planet system GJ\,1214 
\citep{berta:2012}. The law was first proposed in \citet{diaz:1992} and 
describes the specific intensity as

\begin{align}
I(\mu)/I(1) &= 1 - c (1-\mu) - d (1-\sqrt{\mu}),
\label{eqn:sqrtlaw}
\end{align}

where $c$ and $d$ are the two LDCs associated with this law. Following the same 
procedure as used earlier in \S\ref{sub:derivation}, imposing the condition of 
an everywhere-positive profile yields

\begin{align}
c + d < 1.
\end{align}

Similarly, the condition of a monotonically decreasing intensity profile from
centre-to-limb gives two constraints:

\begin{align}
&d > 0,\\
&2 c + d > 0.
\end{align}

These three non-parallel conditions are easily imparted using the triangular 
sampling technique and using the replacements $q_1^{\mathrm{sqrt}}$ and 
$q_2^{\mathrm{sqrt}}$ defined over the interval $[0,1]$:

\begin{align}
q_1^{\mathrm{sqrt}} &= (c + d)^2,\\
q_2^{\mathrm{sqrt}} &= \frac{d}{2(c + d)}.
\end{align}

Alternatively, sampling from uniform, bivariate Dirichlet distribution,
$\mathcal{P}(\balpha=\mathbf{1};v_1^{\mathrm{sqrt}},v_2^{\mathrm{sqrt}})$, may
be achieved using:

\begin{align}
v_1^{\mathrm{sqrt}} &= d/2,\\
v_2^{\mathrm{sqrt}} &= 1 - c - d.
\end{align}

\subsection{Logarithmic law}
\label{sub:loglaw}

\citet{klinglesmith:1970} proposed a logarithmic limb darkening law with
the following form

\begin{align}
I(\mu)/I(1) &= 1 - A (1-\mu) - B \mu (1-\log\mu),
\label{eqn:loglaw}
\end{align}

where $A$ and $B$ are the two associated LDCs. Again, 
following the procedure used earlier in \S\ref{sub:derivation}, we find that
imposing the condition of an everywhere-positive profile yields

\begin{align}
A < 1.
\end{align}

Similarly, the condition of a monotonically decreasing intensity profile from
centre-to-limb gives two constraints:

\begin{align}
&A + B > 0,\nonumber\\
&B < 0.
\end{align}

These three non-parallel conditions are again easily imparted using the 
triangular sampling technique and using the replacements $q_1^{\mathrm{log}}$ 
and $q_2^{\mathrm{log}}$ defined over the interval $[0,1]$:

\begin{align}
q_1^{\mathrm{log}} &= (B + 1)^2,\\
q_2^{\mathrm{log}} &= \frac{A - 1}{B + 1}.
\end{align}

Alternatively, sampling from uniform, bivariate Dirichlet distribution,
$\mathcal{P}(\balpha=\mathbf{1};v_1^{\mathrm{log}},v_2^{\mathrm{log}})$, may
be achieved using

\begin{align}
v_1^{\mathrm{log}} &= 1 - A,\\
v_2^{\mathrm{log}} &= -B.
\end{align}

\subsection{Exponential law}
\label{sub:explaw}

The final two-parameter limb darkening law we consider comes from 
\citet{claret:2003} and takes the form

\begin{align}
I(\mu)/I(1) &= 1 - g (1-\mu) - h \frac{1}{1-e^\mu},
\label{eqn:explaw}
\end{align}

where $g$ and $h$ are the two associated limb darkening coefficients. Following 
the procedure used earlier in \S\ref{sub:derivation} once more, we find that
imposing the condition of an everywhere-positive profile yields two constraints
(unlike all previous examples where this condition only imposed one meaningful
constraint):

\begin{align}
&h < 1 - e^1,\nonumber\\
&h < 0.
\end{align}

However, these two condition are parallel and since $0>(1-e^1)$, then the
two conditions simply boil down to $h<(1-e^1)$. Similarly, the condition of a 
monotonically decreasing intensity profile from centre-to-limb gives two 
constraints:

\begin{align}
&h < 0,\nonumber\\
&\frac{h_1}{h_2} > \frac{e^1}{(1-e^1)^2}.
\end{align}

The first of these two conditions is parallel to the previously derived 
constraint of $h<(1-e^1)$ and in fact less constraining and so we can discard
it. In total then, we have only two non-parallel boundary conditions. As a
result, a triangular enclosed region is not formed in the joint probability 
distribution and so the triangular sampling technique discussed in this paper
is not applicable.

\section{Discussion \& Conclusions}
\label{sec:discussion}

In this paper, we have presented new parametrizations for the LDCs of several 
two-parameter limb darkening laws, including the popular quadratic 
(\S\ref{sec:quadratic}) and square-root laws (\S\ref{sub:sqrtlaw}). When sampled 
over the interval $[0,1]$, our parametrizations exclusively sample the complete 
range of physically plausible LDCs (100\% efficient and 100\% complete). This is 
twice as efficient as the next best parametrization proposed previously 
(\S\ref{sec:comp}). In the case of the quadratic law, we show that our 
parametrization also reduces the mutual correlation between the two LDCs 
(\S\ref{sub:correls}) with a natural geometric explanation (\S\ref{sub:pal}), 
although this was not the motivation behind our formulation.

Fitting astronomical data with our parametrization for the LDCs ensures that all
model parameters fully account for one's ignorance about the stellar intensity
profile, leading to more realistic uncertainty estimates. Derived parameters
make no assumption about the stellar atmosphere model, except the type of
polynomial used to describe it (for which we provide several choices) and that
the observations are of normal, main-sequence stars in broad bandpasses. These
parametrizations are applicable to any observation affected by limb darkening,
such as optical interferometry, microlensing, eclipsing binaries and transiting
planets.

Our parametrization may be explained as follows. Requiring the intensity
profile to be everywhere-positive and monotonically decreasing from 
centre-to-limb imposes three non-parallel boundary conditions on two LDCs (see 
equation~\ref{eqn:conditions}). Given the two LDCs live on a two-dimensional
plane, the three boundary conditions describe a triangular region where 
physically plausible LDCs may reside. This triangular region can be sampled
uniformly by re-parametrizing the LDCs from $\{u_1,u_2\}$ to $\{q_1,q_2\}$ 
(see equation~\ref{eqn:myinverseeqn}) according to a technique used in
computer graphical programming \citep{turk:1990}: triangular sampling. An 
equivalent method is to draw a random variate from a uniform, bivariate
Dirichlet distribution.

We note that the solution is general to any situation where two parameters are 
bound by three non-parallel boundary conditions. Or, even more generally, when 
$N$ parameters are mutually constrained by $N+1$ non-parallel boundary 
conditions leading to tetrahedral sampling and hyper-tetrahedral sampling. In
the case of exoplanet transits, we are therefore faced with the unusual
case of the field of exoplanets drawing from computer games, rather than the
other way around.

\section*{Acknowledgements}

This work was performed [in part] under contract with the California Institute 
of Technology (Caltech) funded by NASA through the Sagan Fellowship Program 
executed by the NASA Exoplanet Science Institute. Thanks to G. Turk \& J. Irwin
for useful discussions and comments in preparing this manuscript. Special
thanks to the anonymous reviewer for his/her positive and constructive feedback.




\label{lastpage}

\end{document}